\documentclass[
aps,%
12pt,%
final,%
notitlepage,%
oneside,%
onecolumn,%
nobibnotes,%
nofootinbib,
superscriptaddress,%
noshowpacs,%
centertags]%
{revtex4}

\begin{document}
%
\title{Transverse momentum spectra of hadrons produced in central heavy-ion 
collisions}
\author{\firstname{G.I.} \surname{Lykasov}}
\affiliation{JINR, Dubna, Moscow region, 141980, Russia}
\author{\firstname{A.N.} \surname{Sissakian}}
\author{\firstname{A.S.} \surname{Sorin}}
\author{\firstname{V.D.} \surname{Toneev}}
\noaffiliation
\begin{abstract}
In-medium effects of transverse-mass distributions of quarks and
gluons are considered assuming a possible local equilibrium  for
colorless quark objects like mesons and baryons created in central
A+A collisions.  It is shown that the average square of the
transverse momentum for these partons grows and then saturates
when the initial energy increases. Within the quark-gluon string
model it leads to the energy dependence of hadron transverse mass
spectra which is similar to that observed in heavy ion collisions.
Comparison with other scenarios is given.
\end{abstract}
\maketitle
\section{\bf Introduction}

Searching for a new physics in heavy-ion collisions at AGS, SPS
and RHIC energies has led to intense theoretical and experimental
activities in this field of research \cite{QM}. In this respect
the search for signals of a possible transition of hadrons into
the QCD predicted phase of deconfined  quarks and gluons,
quark-gluon plasma (QGP), is of particular interest. Among many
signals of the QGP formation, one of the earliest is based on the
relation of the thermodynamic variables, temperature and entropy,
to the observable quantities, the average transverse momentum and
 multiplicity, respectively. Shuryak~\cite{Sh80,ShZ80} and
 van Hove~\cite{vH82} were first to put forth arguments that the
 transverse momentum as a function
 of multiplicity reflects general properties of the equation of state
 of hot and dense matter. In the case of  the first
 order phase transition this relation is quite specific due to
 inevitable formation of the coexistence phase of quark-gluon plasma
 and hadrons, the mixed phase (MP).
 Experimental detection of the QGP phase and the MP in A+A
collisions is a non-trivial task because of smallness of the
space-time volume of the hot and dense  system and possible
contributions of hadronic processes simulating signals of QGP and
MP formation.
Nevertheless,
recent experimental study of the transverse-mass spectra of kaons
from central Au+Au and Pb+Pb collisions revealed "anomalous"
dependence on the incident energy. The effective transverse
temperature (the inverse slope-parameter of the transverse mass
distribution at the mid-rapidity) rather fast increases with
incident energy in the AGS domain~\cite{Ah00}, then saturates at
the SPS energies~\cite{NA49} and approaching the RHIC energy
region~\cite{RHIC} increases again. In agreement with
expectations
this saturation was assumed to
be associated with the deconfinement phase transition and
indication on the MP~\cite{GGB03,Mohanty:2003}. This assumption on
the first order phase transition was indirectly confirmed by the
fact that two independent transport models, based on hadron-string
dynamics, UrQMD and HSD,  failed to reproduce the observed
behavior of the effective kaon
temperature~\cite{Bratk:04,Bratk:05}.
Though the calculated transverse mass spectra are in reasonable agreement
with the experimental results for $pp$, $pA$
and light nuclei (central C+C and Si+Si) collisions, both models strongly 
underestimate the transverse energy $T$ above energy $~5$ AGeV and are not able 
to describe the rapid increase of the inverse slope parameter for collisions
in the AGS domain and its subsequent flattening in the low energy
SPS regime. The failure was attributed to a lack of pressure. This
additional pressure must be generated in the early nonhadronic
phase of the collision because the strong hadronic interactions in
the later stages do not produce it~\cite{Bratk:04,Bratk:05}. The
anomalous effective-temperature behavior was quite successfully
reproduced within hydrodynamic model with equation of state
involving the phase transition~\cite{Braz04}. However, this result
is not very convincing since to fit data the required
incident-energy dependence of the freeze-out temperature should
closely repeat the shape of the corresponding effective kaon
temperature and thereby the problem of the observed anomalous
inverse-slope dependence is readdressed to the problem of the
freeze-out temperature. 

In addition, as was shown in \cite{IR06},
these experimental data within the considered energy range up to
$\sim 160$ AGeV may be reasonably described in the 3-fluid
relativistic hydrodynamic model with the pure hadronic equation of
state. Note that this 3-fluid model also well describes
a large variety of global observables~\cite{IRT06}. This implies
that a hydrodynamic collective motion in the expansion stage,
which is absent in transport codes, plays an important role in the
anomaly discussed. It is of interest that recently a step to taking
into account some collective motion in the microscopic description
has been done in \cite{LBGM07} where a hadronic transport model
has been extended to include three-body collisions. The agreement
between the transport model and experiment for the kaon
transverse-temperature excitation function was noticeably 
improved.

In this paper we propose another way to introduce a collectivity
effect in nuclear system via some  in-medium effect. We consider
the temperature dependence of quark distribution functions inside
a colorless quark-antiquark or quark-diquark system (like meson or
baryon, $h$) created in central A+A collisions. A contribution of
this effect into transverse momentum spectra of hadron is
estimated and it is shown that  it results in larger values of the
inverse slope parameter  and therefore in broadening of the
transverse mass spectra.

\section{\bf Quark in a hadron embedded in equilibrated matter}

Let us assume the local equilibrium  in a fireball of hadrons
whose distribution function can be presented in
the following relativistic invariant form: %
\begin{eqnarray}
f_h^{A}~=~ C_T\left\{1~\pm~exp((p_h\cdot
u-\mu_h)/T)\right\}^{-1}~, \label{def:hdmp} 
\end{eqnarray}
where  $p_h$ is the four-momentum of the hadron, the four-velocity
of the fireball in the proper system is $u=(1,0,0,0)$, the sign
"$+$" is for fermions and "$-$" is for bosons, $\mu_h$ is the
baryon chemical potential of the hadron $h$, $T$ is the local
temperature, and $C_T$ is the $T$-dependent normalization factor.
The distribution function of constituent quarks inside $h$ which
is in local thermodynamic equilibrium with the surrounding nuclear
matter, $f_q^A(x,{\bf p}_t)$, can be calculated using the procedure
suggested for a free hadron in Ref.\cite{Weiskopf:1971}. 
So, the quark distribution in a free hadron moving with the momentum 
$P$ is given by the convolution
 \begin{eqnarray} 
f_{q_v}^h(p_z,{\bf
p}_t)=\int dp_{1z} d^2p_{1t} \ q_v(p_{z},{\bf p}_t) \
q_r(p_{1z},{\bf p}_{1t}) \ \delta(p_z+p_{1z}-P) \
\delta^{(2)}({\bf p}_{1t}+{\bf p}_t) ~, 
\label{def:fqvKW}
 \end{eqnarray}
where $q_v(p_z,{\bf p}_t)$ is related with the probability to find
the valence quark with longitudinal momentum $p_z$ and transverse
momentum ${\bf p}_t$ in the hadron, whereas $q_r(p_{1z}, p_{1t})$
is the probability that all the other hadron constituents (one or
two valence quarks plus any number of quark-antiquark $q{\bar q}$
pairs and gluons) carry the total longitudinal momentum $p_{1z}$ and
the total transverse momentum $p_{1t}$.

Introducing the new variables $x=p_z/P, x_1=p_{1z}/P$, which at
large $P$ become the relativistic invariant light cone variables,
we can rewrite eq.(\ref{def:fqvKW}) (for example, for the
valence quark distribution) in the following equivalent form
\cite{Capella:1981,Benhar:1997}:
 \begin{eqnarray}
f_{q_v}^h(x,p_t)=\int_0^1dx_1\int d^2p_{1t} \ q_v(x,{\bf p}_t) \
q_r(x_1,{\bf p}_{1t}) \ \delta(x+x_1-1) \ \delta^{(2)}({\bf
p}_{1t}+{\bf p}_t) ~, \label{def:fqv}
 \end{eqnarray}

Assuming the factorization hypothesis 
$q_{v,r}(x,{\bf p}_t)=q_{v,r}(x) \ g_{v,r}({\bf p}_t)$ and integrating
eq.(\ref{def:fqv}) with respect to $dx_1$ and $d^2p_{1t}$ we have
 \begin{eqnarray}
f_{q_v}^h(x,p_t)=f_{q_v}^h(x) \ g_{q_v}^h({\bf p}_t)~,
\label{def:fqvfact}
\end{eqnarray}
 where
 \begin{eqnarray} 
f_{q_v}^h(x)=q_{v}(x) \ q_{r}(1-x)~,~ g_{q_v}^h({\bf
p}_t)=g_{v}({\bf p}_t) \ g_{r}(-{\bf p}_t) \label{def:qvgv}
 \end{eqnarray}
For simplicity one can choose the normalized $p_t$-distribution
$g_q^h({\bf p}_t)$ in the Gaussian form
 \begin{eqnarray} 
g_q^h({\bf p}_t)=\frac{\gamma_q}{\pi} \ \exp(-\gamma_q {\bf p}_t^2)~.
\label{def:qptdis} 
\end{eqnarray}
Then we get 
\begin{eqnarray}
g_{v}({\bf p}_t)=g_{r}(-{\bf p}_t)=\left (\frac{\gamma_q}{\pi}\right )^{1/2}
\ \exp(-\frac{\gamma_q}{2} {\bf p}_t^2). 
\label{def:gpt}
\end{eqnarray}

Similarly to eq.(\ref{def:fqvKW}), one can present the quark
distribution inside the hadron $h$ to be in local thermodynamic
equilibrium with the surrounding matter, $f_q^A(x,{\bf p}_t)$, in
the convolution form as
\begin{eqnarray} 
f_q^A(p_{z},p_{t})=\int
dp_{1z}dp_{hz}d^2{\bf p}_{1t}d^2{\bf p}_{ht} \ {\tilde
q}_v(p_{z},{\bf p}_{t}) \ {\tilde q}_r(p_{1z},{\bf p}_{1t}) \
f_h^{A}(p_{ht},{\bf p}_{ht})
\\ \nonumber
\times \delta(p_z+p_{1z}-p_{hz}) \ \delta^{(2)}({\bf  p}_{t}+
{\bf p}_{1t}-{\bf  p}_{ht})~.
\label{def:fqvKWA}
\end{eqnarray}
 In contrast to eq.(\ref{def:fqvKW})  the distribution of the hadron
$h$ in a fireball $f_h^A$ is included in eq.(8), therefore we
integrate over the longitudinal and transverse momenta of $h$. In
the general case, the functions $q_v(p_{iz},{\bf p}_{it})$ and
$q_r(p_{iz},{\bf p}_{it})$ entering into eq.(\ref{def:fqvKW}) may
differ from ${\tilde q}_v(p_{iz},{\bf p}_{it})$  and ${\tilde
q}_r(p_{iz},{\bf p}_{it})$ in eq.(8). Let us introduce the Feynman
variables $x=2p^*_z/\sqrt{s^\prime}, x_1=2p^*_{1z}/\sqrt{s^\prime},
x_h=2p^*_{hz}/\sqrt{s^\prime}$, where $p^*_z,p^*_{1z},p^*_{hz}$
are the longitudinal momenta and 
$s^\prime$ is some characteristic energy squared scale.
Then  eq.(8)
can be reduced to the following form similar to
eq.(\ref{def:fqv}):
\begin{eqnarray} 
f_{q_v}^A(x,p_t)=\int_0^1
dx_1\int_0^1dx_h\int d^2p_{1t}d^2p_{ht} \ {\tilde q}_v(x,{\bf
p}_t) \ {\tilde q}_r(x_1,{\bf p}_{1t})
\\ \nonumber
\times f_h^{A}(x_h,{\bf p}_{ht}) \ \delta(x+x_1-x_h) \
\delta^{(2)}({\bf p}_{1t}+{\bf p}_t-{\bf p}_{ht}) ~.
 \label{def:fqvA}
\end{eqnarray} 
If the factorization hypothesis ${\tilde q}_{v,r}(x,{\bf
p}_t)={\tilde q}_{v,r}(x)\ {\tilde g}_{v,r}(p_t)$ and  the
Gaussian form for ${\tilde g}_{v,r}({\bf p}_{t})$ are assumed,
 \begin{eqnarray}
 {\tilde g}_{x,r}({\bf p}_{t})=\left
(\frac{{\tilde\gamma}_q}{\pi}\right )^{1/2} \
\exp(-\frac{{\tilde\gamma}_q}{2} {\bf p}_t^2)~, 
\label{def:gptA}
\end{eqnarray}
we can get the following expression for $f_{q_v}^A(x,p_t)$
(see {\bf APPENDIX}) :
\begin{eqnarray} 
f_q^{A}(x, {\bf p}_t;T)&=& 2 m_h
TC_T\exp(-\frac{m_h-\mu_h}{T})(1+\frac{T}{m_h})
\frac{1}{\pi}\int_x^1 dx_h \ {\tilde q}_v(x) \ {\tilde q}_r(x_h-x)
\nonumber \\ &\times&
{\tilde\Gamma}_q(x_h)\exp(-{\tilde\Gamma}_q(x_h) p_t^2)~,
\label{def:fqAx}
\end{eqnarray}
 where ${\tilde m}_h(x_h)=\sqrt{m^2_h+x_hs^\prime/4}$, by assumption  
${\tilde\gamma}_q=\gamma_q$ and
\begin{eqnarray} 
{\tilde\Gamma}_q(x_h)=\frac{\gamma_q(1+\gamma_q{\tilde m}_h(x_h)T/2)} 
{1+\gamma_q {\tilde m}_h(x_h)T}~, 
\label{def:tGm}~.
\end{eqnarray}
 From the normalization relation for $f_q^{A}(x, {\bf p}_t;T)$
 \begin{eqnarray}
 \int_0^1 dx\int d^2p_t \ f_q^{A}(x, {\bf p}_t;T)&=&1
\label{def:normfqA} 
\end{eqnarray} 
we can find that $C_T=(2m_h T(1+T/m_h))^{-1}\exp[(m_h-\mu_h)/T]$, ${\tilde q}_v(x)=q_v(x)$ 
and
\begin{eqnarray}
 q_r(1-x)=\int_x^1 dx_h \ {\tilde q}_r(x_h-x)\equiv
\int_0^{1-x}{\tilde q}_r(y)dy~. 
\label{def:tildeqr}
\end{eqnarray}
 Then eq.(\ref{def:fqAx}) reads
\begin{eqnarray} 
f_q^{A}(x, {\bf p}_t;T)&=& \frac{1}{\pi}\int_0^{1-x} dx_1 \
{\tilde q}_v(x) \ {\tilde q}_r(x_1) \
{\tilde\Gamma}_q(x_1+x)\exp(-{\tilde\Gamma}_q(x_1+x) p_t^2)~.
\label{def:fqAxapp}
\end{eqnarray}
 As is evident from eq.(\ref{def:fqAxapp}),
at vanishing temperature $T=0$ the quark distribution $f_q^{A}$
reproduces the quark distribution in a free hadron.

Using the quark distribution $f_{q_v}^A(x,p_t)$ 
we can calculate the average value for the transverse momentum
squared for the quark $<p_{q,t}^2>_h^A$ in $h$ as a function of $x$
and $s_{NN}$:
\begin{eqnarray}
 <p_{q,t}^2(x)>_h^A&=&\frac{\int f_q^A(x,{\bf p}_t)p_t^2d^2p_t}
 {\int f_q^A(x,{\bf p}_t)d^2p_t}~.
\label{def:aptsqA}
\end{eqnarray}
Applying eq.(\ref{def:fqAxapp}), 
 we have
\begin{eqnarray}
 <p_{q,t}^2(x)>_h^A&=&C\int_0^{1-x}dx_1\frac{{\tilde q}_r(x_1)
 (<p_t^2>_q^h+{\tilde m}_h(x_1+x)T)}
{1+{\tilde m}_h(x_1+x)T/(2<p_t^2>_q^h)}~, \label{def:aptsqAG}
\end{eqnarray}
where $<p_t^2>_q^h=1/\gamma_q$ is the average transverse momentum
squared for a quark in the free hadron and
\begin{eqnarray}
C^{-1}&=&\int_0^{1-x}dx_1{\tilde q}_r(x_1) \label{def:mormC}~.
\end{eqnarray}
At $x\simeq 0$ eq.(\ref{def:aptsqAG}) can be presented in the
following equivalent form:
\begin{eqnarray}
 <p_{q,t}^2(x\simeq 0)>_h^A&=&{\tilde C}\int_0^1dx_1
 \frac{{\tilde q}_r(x_1)(<p_t^2>_q^h+(\sqrt{m_h^2+x_1^2s^\prime/4}T)}
{1+(\sqrt{m_h^2+x_1^2s^\prime/4})T/(2<p_t^2>_q^h)}~,
\label{def:aptsqAGt}
\end{eqnarray}
where
\begin{eqnarray}
{\tilde C}^{-1}&=&\int_0^1 dx_1{\tilde q}_r(x_1)~.
\end{eqnarray}
 In the general
case the quark distribution in a nucleon $f_q^N(x)$ at low momentum
transfer when its $Q^2$ QCD evolution can be neglected is
presented as follows:
\begin{eqnarray} 
f_q^N(x)=C_qx^a(1-x)^b~,
\label{def:fqNx}
\end{eqnarray}
 where $C_q$ is the normalization factor. The
parameters $a$ and $b$ can be extracted from the deep inelastic
scattering or calculated within some quark models. Therefore,
according to eq.(\ref{def:tildeqr}), the function
${\tilde q}_r(x_1)$ entering into
eqs.(\ref{def:aptsqAG},\ref{def:mormC}) has the following  form:
\begin{eqnarray}
{\tilde q}_r(x_1)&=&bx_1^{b-1}~. \label{def:qrtilde}
\end{eqnarray}
Usually, $b\ge 1.5$ (see below). So ${\tilde q}_r(x_1)$ falls down
very fast when $x_1$ decreases from $1$ to $0$ and may be taken
out of the integral in eq.(\ref{def:aptsqAGt}) at $x_1=1$ :
 \begin{eqnarray}
<p_{q,t}^2(x\simeq 0)>_{h,appr.}^A&\simeq&\frac{<p_t^2>_q^h+T\sqrt{m_h^2+s^\prime/4}}
{1+T\sqrt{m_h^2+s^\prime \ / \ 4}/(2<p_t^2>_q^h)}~,
\label{def:aptsqappr}
 \end{eqnarray}
 As is evident from eq.(\ref{def:aptsqappr}) the quantity
$<p_{q,t}^2>_h^A$ depends on the energy $\sqrt{s^\prime}$
and temperature $T$. At any nonzero values  of $T$ it grows when
$\sqrt{s^\prime}$ increases (because its derivative with respect to
$\sqrt{s^\prime}$ is positive) and then saturates, reaching the
asymptotic value about $2<p_t^2>_q^h$ at high energies. The
integration over $dx_1$ in eq.(\ref{def:aptsqAGt}) does not change
this result qualitatively.

To calculate  $<p_{q,t}^2(x)>_h^A$ more accurately, we have to know
the functions $q_v(x)={\tilde q}_v(x)$ and $q_r(1-x)$ which is
related to ${\tilde q}_r(y)$, see
eqs.(\ref{def:tildeqr}),(\ref{def:qrtilde}). Parameters of these
distributions can be defined by application of  the quark-gluon
string model (QGSM)~\cite{Kaid1,Kaid2} based on the Regge
asymptotic of quark distributions in a nucleon and $1/N$ expansion
in QCD \cite{tHooft,Veneziano} ($N$ is the number of flavors or
colors). According to this model, the nucleon consists of a
quark, diquark, and quark-antiquark see ($q{\bar q}$). For
example, for the valence $u$-quark in the proton \cite{Kaid2} we
have
\begin{eqnarray} 
a=-\alpha_R(0)~;~b=\alpha_R(0)-2\alpha_N(0)
\end{eqnarray}
where $\alpha_R(0)=1/2$ is the reggeon intercept of the  Regge
trajectory, $\alpha_N(0)=-0.5$ is the intercept of the nucleon
Regge trajectory. Comparing eq.(\ref{def:fqNx}) and
eq.(\ref{def:qvgv}) one can find that
\begin{eqnarray}
u_v(x)=x^{-\alpha_R(0)}~;~u_r(1-x)=(1-x)^{\alpha_R(0)-2\alpha_N(0)}
\label{def:uvur}
\end{eqnarray}
The similar form can be obtained for the $x$-distribution of the valence
$d$-quark in the nucleon. Then, using eq.(\ref{def:aptsqAGt}), the
average transverse momentum  squared can be estimated for
$u$-quark inside the proton which is in local equilibrium in the
fireball $<p_{u,t}^2>_p^A$. This quantity at $x\simeq 0$ is
presented in Fig.1 as a function of $\sqrt{s^\prime}$. 
The behavior of this quantity given by eq.(\ref{def:aptsqAGt}) is similar to that
discussed above in respect to the approximate expression (\ref{def:aptsqappr}).

\section{Transverse momentum spectra of hadrons from
central $A+A$ collisions}

Let us now estimate the $p_t$ distribution of the hadron $h_1$
produced after collision of two hadrons  one of them is locally 
equilibrated in a fireball.
We shall explore the QGSM based on $1/N$ expansion in QCD.
Actually, this is the expansion of the QCD amplitude in different topologies
\cite{tHooft,Veneziano}. \footnote{This model, the Quark-gluon
String Model (QGSM) \cite{Kaid1,Kaid2} or  the Dual Parton Model
(DPM) \cite{Capella:1994} differs from the Lund String Model, see
for example Ref.\cite{Werner}.} The first order term is the
so-called planner graphs corresponding to the one-Reggeon exchange
diagrams in the $t$-channel of the hadronic process. The second
order term of this expansion is the so-called cylinder graphs
related to the one-Pomeron exchange diagrams. The last ones make the 
main contribution to inclusive spectra of particles produced in inelastic 
hadronic processes. In Fig.2 the cylinder graphs for inelastic meson-nucleon 
(left diagram) and nucleon-nucleon (right diagram) inelastic processes are presented,
see also Ref.\cite{Veneziano}. According to this model, the
colorless strings are formed between the antiquark/quark (${\bar q
/q}$)   in the colliding meson and the quark/diquak  ($q/qq$) in the colliding
nucleon (left diagram of Fig.2), then, after their break, $q{\bar
q}$ pairs are created and fragmentate into a hadron $h_1$. The
contribution of the cylinder graph (right diagram of Fig.2)
to the inclusive spectrum $\rho_{h_1}^{A}\equiv
E_{h_1}\frac{d\sigma} {d^3p_{h_1}}$ of the hadron $h_1$  from the
collision of two nucleons can be presented as follows:
\cite{Kaid2,LS1}:
 \begin{eqnarray}
\rho_{h_1}^{A}(x,{\bf p}_t;T)
=\sigma_1 [F_q^{h_1}(x_+,p_t;T) F_{qq}^{h_1}(x_-,{\bf p}_t;T)/F_{qq}(0,p_t;T)+\\
\nonumber
F_{qq}^{h_1}(x_+,p_t;T) F_q^{h_1}(x_-,{\bf p}_t;T)/F_q(0,p_t;T)]
~,
\label{def:rhoh}
 \end{eqnarray}
where $\sigma_1$ is the cross section of the 2-chain production,
corresponding to the $s$-channel discontinuity of the cylinder
(one-Pomeron) graph. It is usually calculated within the quasi-eikonal
approximation \cite{Ter-Mart};
\begin{eqnarray}
 F_{q({qq})}^{h_1}(x_{\pm},{\bf p}_t,T)&=&\sum_{flavors}
\int_{x_{\pm}}^1 dx_1\int d^2p_{1t} \ d^2p_{2t} \
f_{q({qq})}^{A}(x_1,{\bf p}_{1t};T) \
G_{q({qq})}^{h_1}(\frac{x_\pm}{x_1},{\bf p}_{2t}) \nonumber \\
&\times& \delta^{(2)}({\bf p}_{1t}+{\bf p}_{2t}-{\bf p}_t)~.
\label{def:Fxpm} 
\end{eqnarray}
Here $x_\pm=\frac{1}{2}(\sqrt{x_t^2+x^2}\pm x)$ and $x_t=2m_{h_1
t}/\sqrt{s_{hh}}$, where $s^\prime$ has been associated with the
energy squared $s_{hh}$ of colliding hadrons,
$m_{h_1 t}= \sqrt{m_{h_1}^2+p_{1t}^2}$;
$z^{-1}G_{q(qq)}^{h_1}=D_{q(qq)}^{h_1}$ is the fragmentation
function (FF) of the quark $q$ (diquark ${qq}$) into the hadron
$h_1$. Actually, the interaction function
$F_{q(qq)}^{h_1}(x_+,p_t,T)$ corresponds to the fragmentation of
the upper quark/diquark ( see right diagram in Fig.2) into the
hadron $h_1$, whereas $F_q^{h_1}(x_-,p_t,T)$ corresponds to the
fragmentation of the down diquark/quark into $h_1$. We can get a
similar expression for the $p_t$-spectrum of the hadron $h_1$ 
produced in the meson-nucleon collision by replacing the 
diquark $qq$ with the antiquark ${\bar q}$.
To calculate the transverse momentum spectrum of the hadron $h_1$ in
the central rapidity region, eq.(\ref{def:rhoh}),  one needs to
know the $p_t$-dependence of the fragmentation function
$D_q^{h_1}$. We assume the same Gaussian dependence  as
eq.(\ref{def:qptdis}). However, the slope of this $p_t$ dependence
$\gamma_c$ can differ from the slope $\gamma_q$ for the constituent
quark $p_t$ distribution. For the mid rapidity region, $x\simeq
0$, eq.(26) can be rewritten in the following form:
\begin{eqnarray}
\rho_{h_1}^{A}(x\simeq 0,{\bf p}_t;T)
=\sigma_1 [F_q^{h_1}(x\simeq 0,p_t;T)) + F_{qq}^{h_1}((x\simeq 0,p_t;T)]
~,
\label{def:rhohxz}
\end{eqnarray}

 We have the following approximate equation for the averaged transverse momentum squared
of the hadron $h_1$ produced in the interaction of the nucleon with the nucleon locally
equilibrated in a fireball:
 \begin{eqnarray}
<p_{{h_1}t}^2>_{NN,appr.}^{AA}\simeq\frac{{\tilde\Gamma}_q(x\simeq
1)+\gamma_c} {{\tilde\Gamma}_q(x\simeq
1)\gamma_c}=\frac{1}{{\tilde\Gamma}_q(x\simeq 1)}+
\frac{1}{\gamma_c}~. \label{def:spthap}
 \end{eqnarray}
By substitution of ${\tilde\Gamma}_q(x\simeq 1)$ given by
eq.(\ref{def:tGm}) at $x_h=1$ we finally have
 \begin{eqnarray}
<p_{{h_1}t}^2>_{NN,appr.}^{AA}\simeq
\frac{<p_t^2>_q^N+T\sqrt{m_N^2+s_{hh}/4}}{1+T\sqrt{m_N^2+s_{hh}/4} \ / \
(2<p_t^2>_q^N)}+ \frac{<p_t^2>_q^N}{r}~, \label{def:spthapp}
 \end{eqnarray}
where $r=\gamma_c/\gamma_q$.

From comparison of eq.(\ref{def:spthapp}) and
eq.(\ref{def:aptsqappr}) one can see that the energy dependence of
the transverse momentum squared for the produced hadron,
$<p_{{h_1}t}^2>_{NN,appr.}^{AA}$, is qualitatively similar to that
of the quark inside the hadron which is in local equilibrium in the
fireball, $<p_{q,t}^2>_h^A$. At large values of
$\gamma_c>>\gamma_q$, the second term in eq.(\ref{def:spthapp})
can be neglected and $<p_{{h_1},t}^2>_{NN,appr.}^{AA}$ as a
function of $\sqrt{s_{hh}}$ and $T$ is close to
$<p_{q,t}^2>_{h,appr.}^A$ given by  eq.(\ref{def:aptsqappr}). In this
case the hadron spectrum copies the quark spectrum.
In fact, $\gamma_c$ is larger than $\gamma_q$ by
factor $r=3-4$ as follows from inclusive $p_t$-spectra of hadrons
produced in hadronic processes analyzed within the
QGSM~Ref.\cite{LS1}.

\section{Results and discussion}

We estimated the average value of transverse momentum squared for
$K^+$-mesons produced in nucleon-nucleon $<p_{K^+,t}^2>_{NN}^{AA}$
and pion+nucleon $<p_{K^+,t}^2>_{\pi N}^{AA}$ interactions  of two hadrons
one of them is thermodynamically locally equilibrated 
in a fireball created in the central $A+A$ collision. This quantity was estimated 
as a function of $\sqrt{s_{hh}}$ at $T=150 \ $ MeV.   
for two cases when $\gamma_c>>\gamma_q$ and $\gamma_c=3\gamma_q$ \cite {LS1} using
eq.(29) for the inclusive spectrum at $x\simeq 0$. 
It is presented in Fig.3.

As is seen from Fig.3, the results obtained are sensitive to
the mass value of the hadron which is in local equilibrium with
the surrounding nuclear matter at $\sqrt{s_{hh}}\leq 10$ GeV.

Collectivity due to thermal effects  results in growth
of $<p_{K^+,t}^2>_{hh}^{AA}$ with energy for colliding hadrons in
central A-A collisions similar to that for unbound quarks in hot
matter, $<p_{q,t}^2>_h^{A}$.  At larger energies $\sqrt{s_{hh}}$
this quantity saturates. The saturation value for this quantity
depends on the hadronization mechanism of quarks/diquarks to
hadrons. A simple exponential estimate of $p_t$-spectra for
produced hadrons $h_1$ is used  to parameterize experimental data:
 \begin{eqnarray} 
\frac{dN}{dm^2_{h_1
t}dy}|_{y=0}~=~C \ exp(-\frac{m_{h_1 t}}{T^*})~, \label{def:mtsp}
 \end{eqnarray}
where the parameter $T^*$ is extracted from fitting experimental
data. There are data on the $T^*$ values for different hadrons and
different $m_t$ domains: "low $p_t$" when $m_{h_1
t}-m_{h_1}~<0.6$~GeV, and "high $p_t$",
$0.6~<~m_{h_1t}-m_{h_1}~<1.6$~GeV  (see for example
Refs.\cite{Ah00,NA49,RHIC}). At low $p_t$ the $m_t$ spectrum given
by eq.(\ref{def:mtsp}) can be presented in the following
approximate form:
 \begin{eqnarray}
\frac{dN}{dm^2_tdy}|_{y=0}~\simeq~Cexp(-m_{h_1})exp(-\frac{p^2_{h_1t}}{2m_{h_1}
T^*}) \label{def:mptsp}
 \end{eqnarray}
Actually, at small transverse momenta $2m_{h_1} T^*\simeq
<p_{h_1,t}^2>^{AA}$, where $<p_{h_1,t}^2>^{AA}$ is the transverse
momentum squared for the hadron $h_1$ produced in central $A+A$
collisions. For $K$-mesons $<p_{K,t}^2>^{AA}\simeq T^** $
$GeV/c^2$. 
The experimental data on the inverse slope
$T^*$ for $K^+$-mesons produced in central Au+Au (Pb+Pb)
collisions as a function of the incident energy per nucleon
  show \cite{Ah00,NA49}
that $T^*$ grows and saturates later on. 

Our results presented in Fig.3 qualitatively
demonstrate similar behavior for $<p_{K^+,t}^2>_{hh}^{AA}$
as a function of $\sqrt{s_{hh}}$. 
One should emphasize that here $\sqrt{s_{hh}}$ is the energy of a pair of 
colliding hadrons and it is not related directly to the initial energy of 
colliding heavy ions.
These results are only an illustration
of the collective effects assumed. In a real case, such binary interactions
occur between various hadrons in a large range of temperatures.

As was noted in the introductional part, the broadening effect for
$m_t$-spectra of the hadron produced in central A+A collisions
observed at AGS, SPS and RHIC energies can qualitatively be
explained using the assumption of possible creation of the QGP
and a co-existing phase of quarks and
hadrons~\cite{GGB03,Mohanty:2003}. From available hydrodynamic
calculations of the transverse inverse-slope excitation
function~\cite{Braz04,IR06} it is not yet clear whether this
behavior can be considered as a signal of the phase transition into
the QGP. Microscopic transport models~\cite{Bratk:04,Bratk:05}
taking into account formation and decay of strings as well as
the multiple rescatterings of hadrons are definitely not able to
describe these data. An attempt to enhance the rescatterings was
undertaken in Ref.~\cite{BBRS04} by including the Cronin effect.
In the transport approach enhancement of the intrinsic quark
transverse momentum spread $<p_t^2>^A_q$ is simulated by
increasing the average transverse momentum of quarks $<p_t^2>_{q}$
with the number of previous collisions of primary nucleons
$N_{prev}$ as
\begin{eqnarray}
<p_{t}^2>_q^A=<p_t^2>_{q} (1+\alpha N_{prev})~, \label{Cron}
\end{eqnarray}
 where the parameter $\alpha\approx 04$. Now the description of
spectra becomes rather good at the RHIC energies, improves
essentially at the SPS energy of 160 AGeV, but does not show any
significant change at 11 AGeV~\cite{BBRS04}. Consequently, the
"pre-hadronic" Cronin effect, realized via eq.(\ref{Cron}), is not
responsible for the anomalous behavior of kaon slopes around AGS
energies.

 Another scenario of collectivity, color rope formation, was
proposed in~\cite{Knoll:1984}. This color rope model assumes that
in central A+A collisions several strings are produced, some of
them on top of each other. The common chromo-electric field
created by overlapping $K$ single guark/antiquark sources may form
a $K$-fold rope. The spread of hadron transverse mass
distributions resulting from color string-rope decay is defined by
the surface tension parameter $\kappa^K$ which is $\sqrt K$
times larger than the appropriate parameter for the decay of a
single string in $N-N$ collisions,  $\kappa^K=\sqrt K \kappa$. 
In a certain sense, this scenario is opposite to the proposed one:
changing originates not from the initial "pre-hadronic" level, but rather 
comes from the final state as in-medium modification of the string break
 function. 
Similar effect for overlapping strings was
estimated in Ref.\cite{BBRS04}. Only a small increase in the inverse
slope parameter at AGS energies was found because the string
densities are low. At SPS and RHIC energies the model gives
hardening of the spectra by about $15\%$~\cite{BBRS04}.

\section{ Conclusion}

We have found that the quark distribution in a hadron depends on
the fireball temperature $T$. 
At any $T$ the average transverse momentum squared
of a quark grows and then saturates when $\sqrt{s^\prime}$ increases. Numerically this saturation 
property depends on $T$. The modification of of initial quark distributions leads 
 to a similar energy dependence for the average transverse 
momentum squared $<p_{h_1,t}^2>_{hh}^{AA}$ of the hadron $h_1$. 
The saturation property for $<p_{h_1,t}^2>_{hh}^{AA}$ depends also 
on  the temperature $T$ and it is very sensitive to the dynamics of hadronization.
As an example, we  estimated the energy dependence of the inverse slope of transverse mass spectrum
of $K$-mesons produced in the interaction of two hadrons in the fireball created in central A+A collisions. 
It is qualitatively similar to the incident
energy dependence of this quantity observed experimentally. 
We guess that our assumption on the thermodynamical 
equilibrium of hadrons given by eq.(\ref{def:hdmp}) can be applied for heavy nuclei only
and not for the early interaction stage.

From the above discussion we see that the observed anomalous
behavior of the kaon inverse slope in central A+A collisions is
still puzzling. There are several scenarios which can be valid in
various degrees, however, the final consistent solution of this
puzzle is still absent. Its solution can partially  be due to
the proposed thermal mechanism of collectivity of hadrons. 
 For  final decision  this and other effects should be taken into
 in a dynamical transport model. The solution of this
"step-like" puzzle is an important point in the scientific
programs on the future heavy-ion accelerators FAIR~\cite{FAIR} and
NICA~\cite{NICA}

\vspace{0.25cm}
{\bf Acknowledgments}\\
The authors are grateful for very useful discussions with
P.Braun-Munzinger, K.A.Bugaev, W.Cassing, A.V.Efremov,   
M.Gazdzicki, S.B.Gerasimov, M.I.Gorenestein, Yu.B.Ivanov, 
A.B.Kaidalov and O.V.Teryaev. This work was supported in part by RFBR 
Grant N 05-02-17695 and by the special program of the Ministry 
of Education and Science of the Russian Federation (grant
RNP.2.1.1.5409).


%
\begin{center}
{\bf Figure Captions}
\end{center}
Fig.1. The energy dependence of the average transverse
momentum squared for the $u$-quark in a proton in nuclear matter
at temperature $T$.

Fig.2. The cylinder graph for the inelastic meson-nucleon
processes (left) and the cylinder graph for the inelastic
nucleon-nucleon reaction (right) \cite{Veneziano} 

Fig.3. The average transverse momentum squared of the $K^+$-meson produced from the
interaction of two hadrons  one of them is in the equilibrated fireball 
as a function of its energy $\sqrt{s_{hh}}$ at $T=0.15$ GeV. 
Curves $1$ and $2$ correspond to $<p_{K^+,t}^2>_{NN}^{AA}$ and
$<p_{K^+,t}^2>_{\pi N}^{AA}$ respectively when $\gamma_c>>\gamma_q$, whereas  curves 
$3$ and $4$ correspond to the same quantities when $\gamma_c=3\gamma_q$. Line $5$ 
corresponds to the average transverse momentum squared
of $K^+$ produced in free  $p+p$ collisions
$<p_t^2>_{K^+}^{NN}=0.14  \ GeV/c^2$.

\newpage
%
\begin{figure}[htb]
\includegraphics[angle=-90,width=1.0\textwidth]{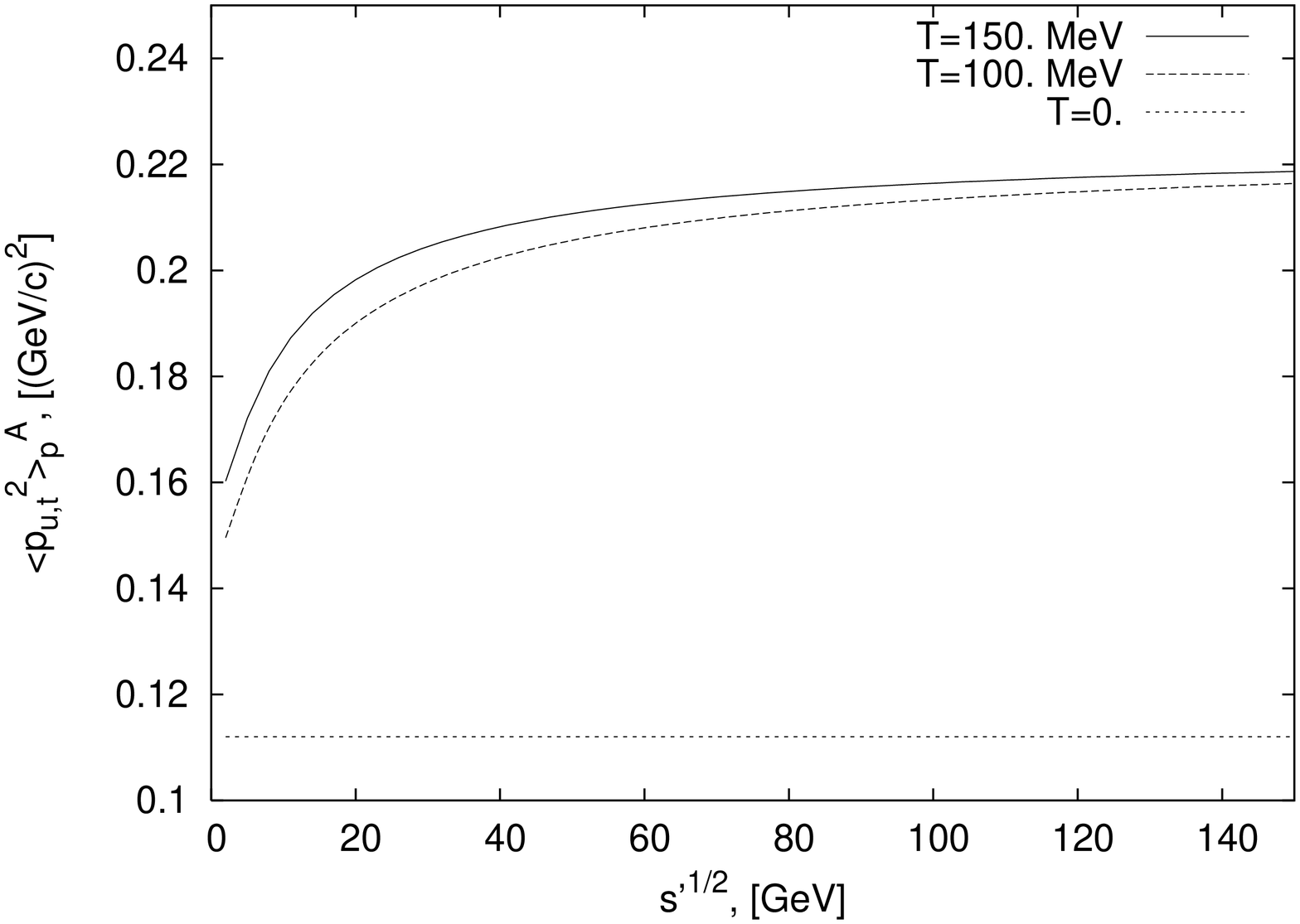}
\caption[Fig.1]{}
\end{figure}

\newpage
%
\begin{figure}
\includegraphics{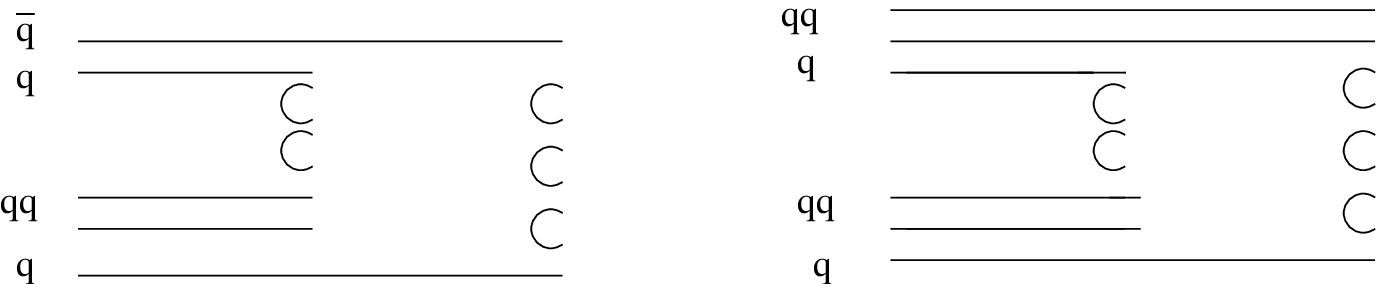}
\caption[Fig.2]{}
\end{figure}

\newpage
%
\begin{figure}[htb]
\includegraphics[angle=-90,width=1.0\textwidth]{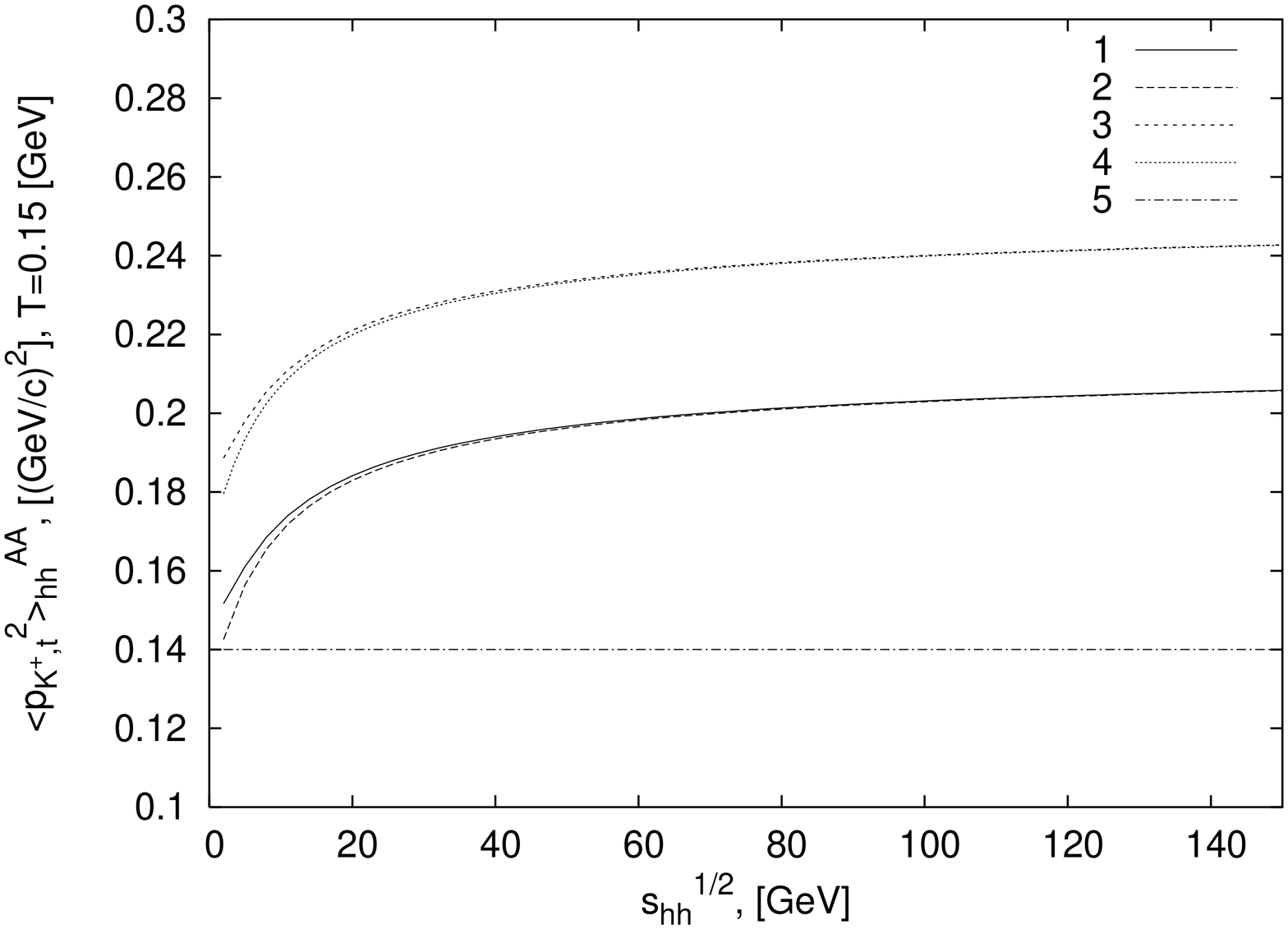}
\caption[Fig.3]{}
\end{figure}


%

\newpage
\begin{center}
{\bf APPENDIX}
\end{center}
Let us get eq.(\ref{def:fqAx}) using eqs.(9,\ref{def:gpt}) and eq.(\ref{def:qptdis}).
Eq.(9) can be also written in the following equivalent form
 \begin{eqnarray} 
f_q^A(x,{\bf p}_t;T)=\frac{\gamma_q}{\pi} C_Texp(\frac{\mu_h}{T})
\int_x^1 dx_h{\tilde q}_v(x){\tilde q}_r(x_h-x)
\exp(-\gamma_q^\prime p_t^2)\times\\ 
\nonumber
\times\frac{1}{(2\pi)^2} \int \exp(-\gamma_q^\prime k_{1t}^2)
\times \exp(-\frac{\sqrt{k_{2t}^2+{\tilde m}_h^2(x_h)}}{T})\exp(i{\vec b}\cdot({\vec k}_{1t}+
{\vec k}_{2t}-{\vec p}_t))d^2k_{1t}d^2k_{2t}d^2b~,
\label{def:apfqHP}
 \end{eqnarray} 
where $\gamma_q^\prime={\tilde\gamma}_q/2,~{\tilde m}_h^2(x_h)=m_h^2+x_h^2s^\prime/4$.
The first integral in eq.(34) reads
 \begin{eqnarray} 
J_1=\int \exp(-\frac{\sqrt{k_{2t}^2+{\tilde m}_h^2(x_h)}}{T})\exp(i{\vec b}\cdot{\vec k}_{2t})d^2k_{2t}\equiv
2\pi \int_0^\infty \exp(-\frac{\sqrt{k_{2t}^2+{\tilde m}_h^2(x_h)}}{T})\times
\\
\nonumber
\times
J_0(bk_{2t})k_{2t}dk_{2t}=2\pi\frac{a_T}{(a_T^2+b^2)^{3/2}}(1+{\tilde m}_h(x_h)\sqrt{a_T^2+b^2})
\exp(-m_h\sqrt{a_T^2+b^2})~,
\label{def:secint}
 \end{eqnarray} 
where $a_T=1/T$, $J_0(bk_{2t})$ is the Bessel function of order 0 depending on $bk_{2t}$ .\\
Eq.(35) in the central region when $b^2<1/T^2$ can be presented in the following approximate
form:
 \begin{eqnarray} 
J_1=2\pi T({\tilde m}_h(x_h)+T)\exp(-b^2{\tilde m}_h(x_h)T/2)~.
\label{def:apsecint}
 \end{eqnarray} 
Using now the form for $J_1$  given by eq.(\ref{def:apsecint})
 we can calculate the second integral in eq.(34)
 \begin{eqnarray} 
\int \exp(-\frac{b^2}{4\gamma_q^\prime})\exp(-\frac{b^2{\tilde m}_hT}{2})exp(-i{\vec b}\cdot{\vec p}_t) d^2b=
 \frac{4\pi\gamma_q^\prime }{(1+2\gamma_q^\prime {\tilde m}_h T)}
\exp(-\frac{\gamma_q^\prime p_t^2}{1+2\gamma_q^\prime {\tilde m}_h T})~.
\label{def:finint}
\end{eqnarray} 
Including all the terms staying in front of eq.(35) we get the form for
$ f_q^A(x,{\bf p}_t;T)$ given by eq.(\ref{def:fqAx}). Note that by getting eq.(\ref{def:fqAx}) the
term
$$
({\tilde m}_h(x_h)+T)\exp(-\frac{{\tilde m}_h(x_h)}{T})
$$
was moved out the integral in eq.(34) at $x_h\simeq 0$ because the exponential function falls down very fast
when $x_h$ increases from $0$ up to $1$.

\end{document}